\documentclass[12pt,preprint]{aastex}
\usepackage{graphicx}
\newcommand\simlt{\lower.5ex\hbox{$\; \buildrel < \over \sim \;$}}
\newcommand\simgt{\lower.5ex\hbox{$\; \buildrel > \over \sim \;$}}

\begin{document}
\title{Observational signatures of sub-photospheric radiation mediated shocks in the prompt phase of GRBs}
\author{Amir Levinson$^{1,2}$}
\altaffiltext{1}{School of Physics \& Astronomy, Tel Aviv University,
Tel Aviv 69978, Israel; Levinson@wise.tau.ac.il}
\altaffiltext{2}{JILA, University of Colorado and National Institute of Standards and Technology, 440 UCB Boulder, Co 80309-0440, USA}

\begin{abstract}
A shock that  form below the photosphere of a GRB outflow is mediated by Compton scattering of radiation advected into the shock by the upstream fluid.  The characteristic scale of such a shock, a few Thomson depths, is larger than any kinetic scale involved by several orders of magnitudes, hence, unlike collisionless shocks,  radiation mediated shocks cannot accelerate particles to non-thermal energies.  The spectrum emitted by a shock that emerges from the photosphere of a GRB jet,
reflects the temperature profile downstream of the shock, with a possible contribution at the highest energies from the shock transition layer itself.  We study the properties of radiation mediated shocks that form during the prompt phase of GRBs, and compute the time integrated spectrum emitted by the shocked fluid following shock breakout. 
We show that the time integrated emission from a single shock exhibits a prominent thermal peak, with the location of the peak depending on the shock velocity profile. 
We also point out that multiple shock emission can produce a spectrum that mimics a Band spectrum.
\end{abstract}

\section{Introduction}
There is mounting evidence that a significant fraction of the bulk energy of GRB outflows is dissipated below the photosphere (e.g., Eichler \& Levinson 2000; Peer, et al. 2006, Ryde \& Peer 2009; Peer, et al. 2012).   Such dissipation may be accomplished through formation of shocks in a hydrodynamic flow (Levinson \& Eichler 1993;  Eichler 1994; Meszaros \& Rees 2000; Giannios 2011), magnetic reconnection in a magnetically dominated flow (Lyutikov \& Blandford 2003; Giannios \& Spruit 2007; McKinney \& Uzdensky 2012), and perhaps other means (e.g., Beloborodov 2010).  Recent simulations of jet breakout from the stellar envelope of a progenitor star indicate that strong tangential collimation shocks form below the photosphere of the expanding outflow, in a region of moderate optical depth ($1<\tau<50$) (Lazzati et al. 2009).  The photosphere is located at a distance of several hundreds stellar radii from the central engine, $r_{ph}\sim 10^{12} - 10^{13}$ cm, with the exact location depending on the type of the progenitor star.  These collimation shocks can dissipate a considerable fraction of the bulk energy, leading to variable emission on intermediate timescales (Morsony et al. 2010).  In addition to the collimation shocks, intermittencies of the central engine are expected to lead to formation of internal shocks, that are commonly thought to be the origin of the short-timescale variability (down to milliseconds) observed in many bursts.  
These shocks can be produced above the photosphere or below it, depending on the bulk Lorentz factor of the expelled outflow and the duty cycle of the central engine (Bromberg et al. 2011, BML11).  It has been shown (BML11) that for moderate bulk Lorentz factors, $\Gamma<500$, and a reasonable duty cycle these shocks should form in a region of moderate optical depth, $1<\tau<10^3$. 

Shocks that traverse an optically thick plasma are mediated by radiation when the shock velocity (measured henceforth in units of c) satisfies 
\begin{equation}
\beta_s >> 4\times10^{-5}n_{15}^{1/6},
\end{equation}
where $n=10^{15}n_{15}$ cm$^{-3}$ is the density of the unshocked plasma (Weaver 1976).   This condition is always satisfied for shocks that form below the photosphere of a GRB outflow during the prompt phase, hence they must be mediated by radiation. In case of a nonrelativistic radiation mediated shock (NRRMS), $\beta_s<<1$, the length scale of the shock transition layer $\Delta^\prime_s$ can be estimated by equating the photon diffusion time across the shock, $t^\prime_D\simeq n_e\sigma_T\Delta_s^{\prime2}/c$, with the shock crossing time, $t^\prime_s=\Delta^\prime_s/(c\beta_s)$.  This yields $\Delta^\prime_s\simeq (n \sigma_T \beta_s)^{-1}$.  In case of a relativistic radiation mediated shock (RRMS) that propagates in a cold medium, the shock width, as measured in the shock frame, is roughly $\Delta^\prime_s\simeq0.1(n\sigma_T)^{-1}\gamma_u^2$ for sufficiently large $\gamma_u$ (Budnik et al. 2010 (BKSW10); Nakar and Sari 2012). 
Here $\gamma_u$ is the Lorentz factor of the upstream fluid with respect to the shock frame.  The width of a RRMS which is dominated by photon advection, as discussed below, may have a different scaling, but is also on the order of $(\sigma_T n)^{-1}$ ( Levinson and Bromberg 2008).  Thus, the width of radiation mediated shocks span a range between a few to a few tens Thomson depths. 

The structure and emission of a RRMS were computed in  BKSW10, assuming that the upstream plasma is cold.  Under this condition the dominant photon source inside and just downstream of the shock transition layer is Bremsstrahlung emission by hot electrons and positrons that are produced via annihilation of counter streaming photons.  Since, as will be shown below,  the photon generation rate is too low to allow for the radiation produced inside the shock to thermalize, the characteristic temperature in the immediate post shock region largely exceeds the black body limit. Typically, $60\ {\rm keV}\simlt T_s\simlt200\ {\rm keV}$ for $\beta_s>0.6$ (Katz et al. 2009 (KBW09)).  The spectrum of the radiation produced inside the shock exhibits a tail that extends to energy in excess of  $\gamma_um_ec^2$ in the shock frame (owing to pair production in the near upstream region).  For typical GRB parameters (as well as for other systems) the length scale of the shock transition layer, $\Delta_s^\prime\sim10^9n_{15}^{-1}$ cm,  is larger by many orders of magnitudes than any kinetic scale involved, and in particular  the plasma skin depth, $c/\omega_p\sim 1 n_{15}^{-1/2}$ cm, and the gyroradius of protons, $r_L\sim3\gamma\beta (B/10^6\ {\rm G})^{-1}$ cm.  An immediate consequence is that radiation mediated shocks cannot accelerate particles, contrary to what is commonly assumed.  Another consequence is that the hot electron-positron pairs are tightly coupled to the protons through electromagnetic interactions and, therefore, thermalize on scales much shorter than the Thomson mean free path.  The local distribution of the electrons and positrons can therefore be taken to be Maxwellian everywhere. 

The conditions invoked in BKSW10 are anticipated in shocks that breakout of the envelope of an exploding star (e.g., Weaver 1976; Chevalier 1992; Rabinak \& Waxman 2011; Nakar \& Sari 2010, 2012;  Sapir et al.  2011), and perhaps in some other situations.  However, as shown in BML11, under the conditions prevail below the photosphere of a GRB outflow, the fluid upstream of a sub-photospheric shock advect photons at a high enough rate to render photon production in the shock transition layer negligible.   As a result, the temperature in the immediate downstream is significantly reduced, but still exceeds the black body limit.   The spectrum emitted by the hot plasma downstream of a shock that breaks out of the photosphere depends on the temperature profile behind the shock.  In what follows, we analyze the properties of sub-photospheric shocks that form in the prompt phase of GRBs,  and compute the spectrum emitted by a single shock that emerges from the photosphere.  In \S 2 we examine the conditions behind a planary shock, both in the relativistic and in the nonrelativistic regimes. In \S 3 we apply the results obtained in \S 2 to analyze the properties of a radiation mediated  shock propagating in a GRB jet, and to compute the emitted spectrum. We conclude in \S 4.

\section{Structure of a radiation mediated shock}
Consider a planar shock propagating in a homogeneous medium consisting of protons, electrons and possibly seed radiation.  Under the conditions envisioned here, the shock is mediated by Compton scattering of radiation produced at about one Thomson depth downstream of the shock transition layer - the deceleration zone where the flow velocity changes  from its far upstream value to its terminal value downstream. 
In the frame of the shock, the jump conditions read:
\begin{eqnarray}
n_{bu}\gamma_u\beta_u=n_{bd}\gamma_d\beta_d,\label{jmp-1}\\
w_u\gamma_u^2\beta_u^2+p_u=w_d\gamma_d^2\beta_d^2+p_d,\label{jmp-2}\\
w_u\gamma_u^2\beta_u=w_d\gamma_d^2\beta_d,\label{jmp-3}
\end{eqnarray}
where subscripts $u$ and $d$ refer to the upstream and downstream values of the fluid parameters, respectively,  $n_b$ denotes the baryon density, $w$ the specific enthalpy, $\beta$ the fluid velocity with respect to the shock frame, and $\gamma=(1-\beta^2)^{-1/2}$ the Lorentz factor.   For the situations considered in this paper the pressure, when important, is always dominated by radiation. Thus,  the specific enthalpy, both upstream and downstream of the shock transition layer, can be approximated as $w=n_{b}m_pc^2+4p_{r}/3$, where $p_r$ denotes the radiation pressure. 

\subsection{RRMS}
We consider first the ultra-relativistic limit of Eqs. (\ref{jmp-1})-(\ref{jmp-3}).
For shocks that form well above the coasting radius of a GRB jet the ratio of radiation pressure and rest mass energy density upstream of the shock, $\tilde{p}\equiv p_{ru}/(n_{bu}m_pc^2)$, is always small, as implied by Eq. (\ref{tild-p}) below.  Thus, to a good approximation we can neglect the pressure of the upstream fluid; that is, we set $p_u=0$ in Eqs. (\ref{jmp-2}) and (\ref{jmp-3}).  Then, the jump conditions, augmented by the equation of state $w_d=n_{bd}m_pc^2+4p_{rd}/3$ downstream, reduce to 
\begin{eqnarray}
\beta_d=1/3,\label{betadR}\\
e_{rd}=2 n_{bu}\gamma_u^2\beta_u^2m_pc^2.\label{erR}
\end{eqnarray}
Now, since the upstream flow is supersonic, the photon density there, $n_{ru}$, is always smaller than that required to establish a full thermodynamic equilibrium downstream.  Therefore, photons must be produced inside and downstream of the shock transition layer in order for the radiation to reach thermal equilibrium.  For the fast shocks considered here the thermalization time is much longer than the shock crossing time (see below).  Consequently, the temperature $T_s$ in the immediate post shock region, where the shock velocity approaches its terminal value $\beta_d$, is well in excess of the black body temperature $T_d$.  Henceforth, we denote quantities associated with the radiation field in the immediate post shock region by subscript $s$, and quantities in the far downstream region by subscript $d$.  The energy density of the radiation in the immediate post shock region, $e_{rs}$, is then given by Eq. (\ref{erR}).  For the infinite planar shock studied here the energy downstream is conserved, so that  $e_{rd}=e_{rs}$. However, under the conditions considered in the following section, where the shock propagates in an inhomogenous, relativistically expanding flow, the radiation cools adiabatically during the evolution of the shock and $e_{rd}$ decreases with time.

The thermalization length depends on the rate at which photons are  generated.  For the situations addressed here the relevant photon sources downstream are free-free emission and double Compton (DC) scattering (Svensson 1984).\footnote{Synchrotron emission is another potential source of photons. However, the characteristic frequency of cyclotron photons emitted by thermal electrons is $h\nu_{syn}=0.03 B_6$ eV, where $B_6=B/(10^6\ G)\sim1$ is roughly the equipartition magnetic field strength of a typical GRB near the photosphere.   It can be readily shown that such soft photons are self-absorbed well before their energy is increased significantly, and therefore cannot contribute to the thermal peak. }
The corresponding photon production rates per unit volume are denoted by $\dot{n}_{ff}$ and $\dot{n}_{DC}$, respectively, and are given explicitly in appendix \ref{sec:tmprof}.
Sufficiently far downstream the radiation eventually thermalizes. Then, the photon density reaches the black body limit, and is given by $n_{rd}=e_{rd}/kT_d$ with $e_{rd}=aT_d^4$.  Using Eq. (\ref{erR}) one finds
\begin{eqnarray}
n_{rd}=2.8\times10^{21}n_{bu15}^{3/4}(\gamma_u\beta_u)^{3/2}\quad {\rm cm^{-3}},\label{nrd}\\
kT_d=0.4 n_{bu15}^{1/4}(\gamma_u\beta_u)^{1/2}\quad {\rm keV},\label{kTd}
\end{eqnarray}
where $n_{bu}=10^{15}n_{bu15}$ cm$^{-3}$.  To get a rough estimate of the thermalization length, we consider each process separately.  When double Compton dominates, the thermalization time is $t^\prime_{DC}=n_{rd}/\dot{n}_{DC}$.  The thermalization length can be expressed as $L^\prime_{DC}=\beta_dct^\prime_{DC}$. When measured in units of the Thomson length it reads
\begin{equation}
\tau^\prime_{DC}=\sigma_Tn_{bd}L^\prime_{DC}\simeq 10^6\Lambda_{DC1}^{-1}n_{bu15}^{-1/2}\gamma_u^{-1}\label{LDC-R},
\end{equation}
where Eq. (\ref{rate-DC}) has been employed with $\Lambda_{DC}=10\Lambda_{DC1}$ and $T=T_d$.  Likewise, for thermalization by free-free emission we obtain from Eq. (\ref{rate-ff}) with $\Lambda_{ff}=10\Lambda_{ff1}$ and Eq. (\ref{nrd}),
\begin{equation}
\tau^\prime_{ff}=\sigma_Tn_{bd}L^\prime_{ff}=1.5\times10^5\Lambda_{ff1}^{-1}n_{bu15}^{-1/8}\gamma_u^{3/4}\label{Lff-R}.
\end{equation}
These rough estimates are in good agreement with the detailed calculations of the temperature profile outlined in appendix \ref{sec:tmprof} (see figure \ref{fig:tmprof}).  Evidently, the length scale over which thermodynamic equilibrium is established largely exceeds the shock width. 

The temperature in the immediate post shock region, at a few Thomson depths downstream of the shock transition, depends on the seed photon density. In what follows we focus on advection dominated shocks for which photon generation inside the shock (but not in the thermalization layer) can be neglected.  Then, the photon density just downstream of the shock transition layer is the advected density compressed by the shock: $n_{rs}=\beta_u\gamma_un_{ru}/(\beta_d\gamma_d)\simeq\sqrt{8}\gamma_u\beta_un_{ru}$. 
Thermal Comptonization of the advected photons by the hot electrons in the immediate downstream results in a Wien spectrum with an average photon energy $<h\nu>=3kT_s=e_{rs}/n_{rs}=e_{rs}/(\sqrt{8}\gamma_u\beta_un_{ru})$ (BKSW10, BML11). By employing (\ref{betadR}) and (\ref{erR}) we obtain
\begin{equation}
kT_s\simeq23\tilde{n}_4^{-1}\gamma_{u}\beta_u\quad {\rm  keV},\label{kTs-R}
\end{equation}
here $\tilde{n}\equiv n_{ru}/n_{bu}=10^4 \tilde{n}_4$ is roughly the dimensionless entropy per baryon carried by the upstream flow. 

In the thermalization layer the temperature changes from $T_s$ to $T_d<T_s$, with $T_d$ given by Eq. (\ref{kTd}).   The temperature profile is determined by the rate at which photons are generated, $dn_r/dt=\dot{n}_{ff}+\dot{n}_{DC}$, subject to $n_r=e_{rs}/(3kT)$, with $e_{rs}$ given by the jump conditions.  The equation governing the change in temperature across the thermalization layer is derived in appendix \ref{sec:tmprof} (Eq. [\ref{tmprof-planar-eq}]). To obtain the temperature profile we numerically integrated Eq. (\ref{tmprof-planar-eq}) from $T=T_s$ at $\tau_d=0$ up to $T=T_d$ at some distance $\tau_d=\tau_{d0}$, after which the temperature remains constant at $T_d$.  Here $\tau_d$ is the distance from the shock front in units of the Thomson length. Examples are shown in figure \ref{fig:tmprof} for upstream density $n_{bu}=10^{15}$ cm$^{-3}$ (corresponding to $\eta=300$ in (\ref{AR-2}) and (\ref{kR-2})) and different values of $\gamma_u$.  When the shock propagates in an expanding medium, as considered in \S 3, adiabatic cooling must be taken into account.  The temperature then obeys Eq. (\ref{tmprof-diff-eq}), which is a generalization of (\ref{tmprof-planar-eq}).

\subsection{\label{sec:NRRMS} NRRMS}
In the non-relativistic limit the jump conditions (\ref{jmp-1})-(\ref{jmp-3}) reduce to
\begin{eqnarray}
\beta_d=\frac{\beta_u}{7}(1+8\tilde{p}/\beta_u^{2}),\label{betadNR}\\
e_{rs}=3p_{rs}=\frac{18}{7} n_{bu}\beta_u^2m_pc^2(1-\tilde{p}/6\beta_u^2),\label{erNR}
\end{eqnarray}
where  $\tilde{p}=p_{ru}/(n_{bu}m_pc^2)$ as before. It is readily seen that a shock can form provided $\beta_u^2>4\tilde{p}/3$. At lower velocities the upstream flow is subsonic.   The advected photon density in the immediate downstream is given by $n_{rs}=\beta_u n_{ru}/\beta_d=7(1+8\tilde{p}\beta_u^{-2})^{-1}n_{ru}$. Substituting the latter result into (\ref{erNR}) we obtain the temperature in the immediate downstream:
\begin{equation}
kT_s=e_{rs}/n_{rs}\simeq10\tilde{n}_4^{-1}\beta_{u}^2(1+8\tilde{p}\beta_{u}^{-2})\quad {\rm keV}.\label{kTs-NR}
\end{equation}
As noted above, the requirement that the flow is supersonic implies $8\tilde{p}\beta_u^{-2}<6$, so that $kT_s<70\tilde{n}_4^{-1}\beta_u^2$ keV.  For the black body temperature we have
\begin{equation}
kT_d=0.43n_{bu15}^{1/4}\beta_{u}^{1/2}\quad {\rm keV}.\label{kTd-NR}
\end{equation}
The thermalization length is obtained as before.  For free-free emission it is 
\begin{equation}
\tau^\prime_{ff}=2\times10^4\Lambda_{ff1}^{-1}n_{bu15}^{-1/8}\beta_u^{11/4},\label{Lff-NR}
\end{equation}
and for DC 
\begin{equation}
\tau^\prime_{DC}=5.4\times10^5\Lambda_{DC1}^{-1}n_{bu15}^{-1/2}(1+8\tilde{p}\beta_u^{-2}).\label{LDC-NR}
\end{equation}

\section{\label{sec:RMS-in-GRBs}Sub-photospheric shocks in GRBs}
We now use the above results to calculate the temperature downstream of a shock that breaks out of the photosphere of a GRB outflow.  
Consider a conical fireball having an isotropic equivalent
luminosity $L_{iso}=10^{52}L_{52}$ ergs s$^{-1}$.   We assume that the fireball is  ejected
with an initial Lorentz factor $\Gamma_0\sim1$  from a compact central engine  of radius
$R_0=10^6R_6$ cm, and that it carries baryons with an isotropic mass loss rate $\dot M_{iso}$.   The properties
of the fireball, and in particular the location of the photosphere, depend of the dimensionless enthalpy
$\eta=L_{iso}/(\dot M_{iso} c^2)$.
When $\eta<\eta_c$, where
\begin{equation}
\eta_c=\left(\frac{\sigma_TL_{iso}\Gamma_0}{4\pi
R_0m_bc^3}\right)^{1/4}=1.8\times10^3L_{52}^{1/4}R_6^{-1/4}\Gamma_0^{1/4},\label{eta_c-def1}
\end{equation}
the fireball is sufficiently opaque, such that the radiation is trapped during the entire acceleration phase.
The major fraction of the explosion energy is
then converted into bulk kinetic energy of the baryons, and the fireball reaches a terminal Lorentz factor $\Gamma_\infty\simeq\eta$ at some
 radius $r_{coast}\simeq\eta R_0/\Gamma_0$, beyond which it continues to coast.
The photosphere is located somewhere in the coasting region, at $r_{ph}>r_{coast}$.
On the other hand, when $\eta>\eta_c$  the fireball will become transparent already during the acceleration phase, before reaching the coasting
radius $r=\eta_c R_0/\Gamma_0$.   The Lorentz factor in that case  may be
close to $\eta_c$ (Nakar et al. 2005). 

We consider situations wherein $\eta<\eta_c$. In that case the optical depth in the coasting region, at some radius $r>r_{coast}$,
can be expressed in terms of the fireball parameters as
\begin{equation}
\tau(r)=\int_{r}^\infty\sigma_T n_b\Gamma^{-1}dr=\frac{\eta_c^4R_0}{\eta^3\Gamma_0 r} .\label{tau-GRB}
\end{equation}
The photospheric radius $r_{ph}$ can be found from the condition $\tau(r_{ph})=1$, and it is readily seen that $r_{ph}=(\eta_c/\eta)^4r_{coast}$.  Also,
at the coasting radius $\tau(r_{coast})=(\eta_c/\eta)^4$.  For the Lorentz factors inferred from observations, $\eta>0.1\eta_c$, the optical depth above the coasting radius, where shocks are likely to form, satisfies $\tau\simlt10^4$.

The above results hold for a conical flow.  Their generalization to a collimated flow is derived in appendix \ref{app:cond-coll}, where it is found that  
for a flow geometry characterized by a cross-sectional radius $a(z)\propto z^{\kappa}$,
$z_{ph}/z_{coast}=(\tilde{\eta}_c/\eta)^{(3+1/\kappa)}$, with $\tilde{\eta}_c=\Gamma_0(\eta_c/\Gamma_0)^{4\kappa/(1+3\kappa)}$.  For $\kappa<1$ we have $\tilde{\eta}_c<\eta_c$, and the condition that the photosphere will be located in the coasting region occurs at lower values of $\eta$.

Now, suppose that the internal energy of the outflow at the coasting radius is dominated by radiation, as expected for hydrodynamic fireballs.  Then, if the outflow remains adiabatic during the coasting phase, the internal energy changes as $e_{r}\propto n_b^{4/3}$, and the photon density as  $n_r\propto n_b$.  
Consequently, the ratio $\tilde{n}=n_r/n_b$ is conserved for an adiabatic flow.  From (\ref{tau-GRB}) one also obtains for the energy density of the radiation near the photosphere, $e_r(r_{ph})=(\eta/\eta_c)^{32/3}e_r(r_{coast})$, and for the flux $r_{ph}^2e_r(r_{ph})\propto(\eta/\eta_c)^{8/3}r_{coast}^2e_r(r_{coast})$.  So unless $\eta\simeq\eta_c$, or a considerable fraction of the bulk energy dissipates near the photosphere, the radiative efficiency of the fireball is extremely small.

To estimate the photon density near the photosphere we suppose that the GRB outflow is adiabatic from its injection point at $r=R_0$ up to the sub-photospheric region where shocks form.  Since, as shown above, $\tilde{n}$ is conserved along adiabatic streamlines, its value near the photosphere equals to its value at the injection point.  From the hydrodynamic equations it can be readily shown that $h\Gamma$ is conserved along streamlines, where $h=1+4p_r/(n_bm_pc^2)$ is the dimensionless enthalpy per baryon.  Thus, $h_0\Gamma_0=\eta$, and since to a good approximation 
$h_0=4p_{r0}/(n_{b0}m_pc^2)$ we have \footnote{At the base of the flow, where the temperature exceeds a few MeV, the radiation is in thermodynamic equilibrium with the $e^\pm$ pairs, and the pressure is $p_0=11aT_0^4/12$.}
\begin{equation}
\tilde{n}=\frac{3\eta}{11\Gamma_0}\left(\frac{m_p c^2}{kT_0}\right).\label{N-1}
\end{equation}
The temperature $T_0$ can be found using energy conservation: $(11aT_0^4/3)\Gamma_0^2\beta_0c4\pi R_0^2=L_{iso}$, and we adopt $\beta_0=1$.  Substituting $T_0$ thereby obtained into Eq. (\ref{N-1}) gives
\begin{equation}
\tilde{n}= 1.8\times10^5(\eta/\eta_c)R_6^{1/4}\Gamma_0^{-1/4}.\label{N}
\end{equation}
It is worth noting that the same result can be obtained upon assuming that at the coasting radius half of the outflow energy is carried by radiation.  We emphasize that the dimensionless entropy given in Eq. (\ref{N}) depends only on the total power and baryon load of the fireball, and not its structure.  It therefore holds for any outflow geometry.

The ratio of photon pressure and rest mass energy density at a given optical depth $\tau>1$ can be obtained using (\ref{tau-GRB}) and (\ref{N-1}) with $\Gamma_0 T_0=\eta T_{coast}$:
\begin{equation}
\tilde{p}(\tau)\equiv\frac{p_r(r_\tau)}{n_b(r_\tau)m_pc^2}=\frac{p_r(r_{coast})}{n_b(r_{coast})m_pc^2}\left(\frac{r_{coast}}{r_\tau}\right)^{2/3}
\simeq(\tau)^{2/3}(\eta/\eta_c)^{8/3},\label{tild-p}
\end{equation}
where $r_\tau=\tau^{-1} (\eta_c/\eta)^4 r_{coast}$ is the radius at which the optical depth equals $\tau$.  Thus, for $\eta<<\eta_c$, $\tilde{p}<<1$ near the photosphere.  The corresponding temperature is readily found from Eq. (\ref{tild-p}): $kT(\tau)=2.5\tau^{1/6}(\eta/\eta_c)^{5/3}(\Gamma_0/R_6)^{1/4}$ keV.

Consider now a RRMS forming in the coasting region, at a radius $r_0<r_{ph}$, where $r_{ph}$ is the photospheric radius, .  As mentioned above, in its rest frame the shock  has a characteristic width $\Delta^\prime_s\simeq(\sigma_Tn_{bu})^{-1}$, where 
$n_{bu}$ is the proper density of the unshocked gas.  In the Lab frame the shock width is given by
\begin{equation}
\frac{\Delta_s}{r_{ph}}=\frac{\Delta_s^\prime}{\eta r_{ph}}=\eta^{-2}(r/r_{ph})^2=(\eta^2\tau^2)^{-1},
\end{equation}
where $\tau=\sigma_Tn_{bu}r/\eta=(r_{ph}/r)$.   Evidently, the shock transition layer broadens as the shock propagates outwards.  Since in the frame of the downstream fluid the shock velocity is $\beta_d=1/3$, the comoving length of the shocked layer (i.e., the downstream region) at some radius $r$ is $\Delta^\prime_d\simeq r/(2\eta)$, and its optical depth is $\Delta\tau_d=\sigma_T n_{bd}\Delta^\prime_d=\sqrt{2}\gamma_u\tau$.\footnote{Note that the optical depth $\tau$ is defined in terms of the density of unshocked fluid, $n_{bu}$, hence the factor $\sqrt{8}\gamma_u$.}  Consequently, the shocked layer becomes progressively more transparent as the shock evolves. The fraction of the dissipation energy contained inside the shock at a radius $r$  is roughly $\Delta\tau_s/\Delta\tau_d\sim (\sqrt{2}\gamma_u\tau)^{-1}$, which can be significant near the photosphere.  From Eq.  (\ref{Lff-R}) we have 
\begin{equation}
\frac{\Delta\tau_d}{\tau_{ff}^\prime}\simeq10^{-5}(\eta/\eta_c)^{1/2}\tau^{5/4}\gamma_u^{-3/4}(R_6/\Gamma_0)^{-1/8}\Lambda_{ff1},\label{therma-ff-dyn}
\end{equation}
and from Eq.  (\ref{LDC-R}) 
\begin{equation}
\frac{\Delta\tau_d}{\tau_{DC}^\prime}\simeq2\times10^{-5}(\eta/\eta_c)^{2}\tau^{2}\gamma_u(R_6/\Gamma_0)^{-1/2}\Lambda_{DC1}.\label{therma-DC-dyn}
\end{equation}
Consequently, thermalization can occur only for shocks that form at small radii, $r/r_{ph}\simlt10^{-4}$.  Since $r_{coast}=(\eta/\eta_c)^4r_{ph}$, the latter condition requires $\eta/\eta_c<0.1$ for shocks that form in the coasting region, at $r>r_{coast}$. 

The observed temperature of a shell that just crossed the shock is found from Eqs. (\ref{kTs-R}) and (\ref{N}):
\begin{equation}
kT_{s,ob}=\eta kT_s\simeq 2 \gamma_u\beta_uL_{52}^{1/4} \Gamma_0^{1/2}R_6^{-1/2}\quad {\rm MeV},\label{kTsR}
\end{equation}
and it is seen that it is independent of the bulk Lorentz factor.  This temperature may be different for different shells if $\gamma_u$ changes during shock propagation. 
The corresponding black body temperature, defined as $T_d=(e_{rs}/a)^{1/4}$, can be expressed as 
\begin{equation}
kT_{d,ob}=\eta kT_d\simeq   1.3\times10^{-3}\eta^2L_{52}^{-1/4}(\gamma_u\beta_u)^{1/2}\quad {\rm keV} .\label{kTdR}
\end{equation}
The maximum observed frequency behind the shock is $h\nu_{max}=3kT_{s,ob}\simeq6\gamma_u$ MeV for our fiducial model.  

The above considerations suggest that the evolution of a shock that propagates in an inhomogeneous, relativistically expanding medium is far more complex than that of an infinite planar shock.  Firstly, the shock velocity $\gamma_u\beta_u$ may not be constant during shock propagation. 
Secondly, as shown above, the shocked shell becomes more transparent as the shock moves outwards. 
This means that photons that were trapped in downstream shells that passed the shock at small radii, can re-interact with the shock at larger radii.  As a consequence, the evolution of shocked shells may not be adiabatic.   Thirdly, the shock width increases, suggesting that near the photosphere a significant fraction of the dissipation energy may be contained inside the shock, which may have important implications for the observed spectrum.  In fact, it is likely that some shocked shells that regain transparency will become part of the shock transition layer.  Nonetheless, from (\ref{therma-ff-dyn}) and (\ref{therma-DC-dyn}) it is evident that photon generation occurs predominantly at small radii (large optical depths), on scales over which density variations and adiabatic cooling are small, so that the number of photons in the shell can be computed to a good approximation using the planary shock model.  The increase in photon density by the thermalization process implies that even if the shocked shells are re-heated at larger radii, the average photon energy will be smaller than $3kT_{s}$ given by Eq. (\ref{kTsR}), as the energy is now shared by more photons.  A complete treatment must account for the temporal evolution of the shock structure and the re-heating of the shocked shells in a self-consistent manner.  Such an analysis is beyond the scope of this paper.  Below, we present sample spectra computed under simplifying assumptions that will be outlined in the next section.

Similar conclusions are obtained for NRRMS. From Eqs. (\ref{kTs-NR}), (\ref{kTd-NR}) and (\ref{N}) we find
\begin{eqnarray}
kT_{s,ob}=\eta kT_s\simeq 1 \beta^2_uL_{52}^{1/4} \Gamma_0^{1/2}R_6^{-1/2}(1+8\tilde{p}\beta_u^{-2})\quad {\rm MeV}.\label{kTsNR}\\
kT_{d,ob}=\eta kT_d\simeq   1.6\times10^{-4}\eta^2L_{52}^{-1/4}\beta_u^{1/2}\quad {\rm keV} .\label{kTdNR}
\end{eqnarray}
The thermalization depth is found to be smaller by a factor of about 10 to 100 than that of RRMS, depending on the value of $\beta_u$.

\subsection{The spectrum emitted by a single shock}
To illustrate some properties of the emitted spectrum, we shall (i) assume  that the shock Lorentz $\gamma_u$ is constant during the evolution, with the exception of one example (figure \ref{fig:spec4}); (ii) ignore possible reheating of the shocked shells near the photosphere; (iii) ignore the contribution of the shock transition layer (which might be important) to the observed spectrum.  As explained above, the spectrum inside the shock is very hard, extending up to about $\gamma_u m_ec^2$ ($\eta\gamma_um_ec^2$ in the observer frame).  So, our analysis below does not account for the portion of the spectrum above $kT_{s,obs}$.

Once the shock breaks out of the photosphere, an observer will start receiving radiation from the shocked shells. Over time deeper and deeper shells will reach the photosphere and emit.  Let $t=0$ be the Lab time at which the shock reaches the photosphere, and let $x_s$ denotes the distance of some shocked shell from the photosphere at $t=0$.  The radius at which that shell was formed (i.e., crossed the shock) is $r_s=r_{ph}-2\eta^2 x_s$, and the corresponding optical depth is $\tau_s$.  The shell will reach the photosphere at time $t=x_s/c$, with a temperature  $T_{ph}(\tau_s)$ given as a solution of Eq. (\ref{tmprof-diff-eq}).  The radiation trapped in the shell has a Wien spectrum.  At the photosphere,
\begin{equation}
I_\nu(t=x_s/c)=\frac{e_r(r_{ph})c}{24\pi}\left(\frac{h}{kT_{ph}(\tau_s)}\right)^4\nu^3e^{-h\nu/kT_{ph}(\tau_s)},\label{I-Wien}
\end{equation}
where $e_r(r_{ph})=2\gamma_u^2n_{bu}(r_{ph})m_pc^2(r_s/r_{ph})^{\alpha/3}=2\gamma_u^2n_{bu}(r_{ph})m_pc^2\tau_s^{-\alpha/3}$ (see Eq. (\ref{app-er-addib})), and we invoke a density profile $n_u\propto r^{-\alpha}$, $\alpha>1$, for the unshocked flow. 
The isotropic equivalent spectral luminosity emitted from the photosphere is given by $L_\nu(t,r_{ph})=4\pi^2 r_{ph}^2 I_\nu(t=x_s/c)$, and the time integrated spectral energy distribution (SED) by
\begin{equation}
\nu E_\nu\equiv\int \nu L_\nu(t,r_{ph})dt\propto   \int_1^{\tau_{0}}\left(\frac{h\nu}{kT_{ph}(\tau_s)}\right)^4\tau_s^{-4\alpha/3(\alpha-1)}e^{-h\nu/kT_{ph}(\tau_s)}d\tau_s.\label{specL}
\end{equation}
Here $\tau_{0}$ corresponds to the radius  $r_0$ at which the shock was initially formed. 

Figures \ref{fig:spec1} - \ref{fig:spec3} display the time integrated SED, $\nu E_\nu$, for a conical flow ($\alpha=2$), and different values of the remaining parameters, assuming $\gamma_u=$ const.  
All curves are normalized to their peak values.  
As seen, the integrated emission exhibits a roughly thermal spectrum, with a peak energy  $h\nu_{peak}\simeq 3kT_{s,ob}$, where $T_{s,ob}$ given by Eq. (\ref{kTsR}) for RRMS and by Eq. (\ref{kTsNR}) for NRRMS. The conceivable effect of reheating near the photosphere and the contribution of the shock transition layer itself are ignored in these examples.  The latter is naively expected to lead to a relatively hard spectral component above the peak energy $h\nu_{peak}=3kT_{s,ob}$, up to about $h\nu\sim \eta\gamma_um_ec^2$.  

The effect of adiabatic cooling is merely to change the slope of the spectrum below the peak, as seen in figure \ref{fig:spec3}.  For a shock forming at an optical depth $\tau_0>>1$, the time integrated SED at energies between  $h\nu_{peak}/\tau_0^{\alpha/3}$ and $h\nu_{peak}$ is a power law with a slope $1+3/\alpha$ (i.e., $\nu E_\nu\propto\nu^{(1+3/\alpha)}$).
The latter result can be derived using the following heuristic argument:  The energy emitted from the photosphere scales as $\nu L_\nu(t,r_{ph})\propto(r_s/r_{ph})^{\alpha/3}$, and the time integrated SED as  $\nu E_\nu\simeq r_s[\nu L_\nu(t,r_{ph})]\propto  r_s^{(1+\alpha/3)}$.  The peak energy of the spectrum emitted from the photosphere at time $t=x_s/c$ scales as $h\nu=3kT_{ph}(\tau_s)\propto (r_s/r_{ph})^{\alpha/3}$.  Thus, $\nu E_\nu\propto\nu^{1+3/\alpha}$.   

The weak dependence of the emitted spectrum on the optical depth $\tau_0$ is a consequence of the fact that the dominant contribution to the time integrated emission comes form regions near the photosphere.  The reason is that our assumption of constant shock velocity,  $\gamma_u\beta_u=$const, implies uniform dissipation, viz., $dU_{diss}=r_s^\alpha e_r(r_s)dr_s\propto dm$, where $dm\propto n_b(r)r^\alpha dr=n_b(r_{ph})r_{ph}^\alpha dr$ is the mass of a shocked shell.   Now, the optical depth across a shell of mass $dm$ scales as $d\tau\propto n_b(r)dr\propto r^{-\alpha} dr$.   Thus, the mass enclosed below the photosphere satisfies $m(\tau)\propto 1-\tau^{-1/(\alpha-1)}$, and it is seen that it is dominated by shells of moderate optical depth.   Because the power emitted from each shell equals the dissipated power $dU_{diss}$ modified by adiabatic cooling, it is clear the integrated emission is dominated by shells created just below the photosphere.  This is the reason for the relatively small effect of adiabatic cooling seen in figure \ref{fig:spec3}.  From the above it is also evident that at most a fraction $(\tau_{th}-1)^{-1}$ of the dissipated energy can thermalize, where $\tau_{th}>>1$ is the thermalization depth.  Consequently, thermalization is ineffective in case of a uniform dissipation profile.   This statement is true in general, regardless of the specific dissipation mechanism. 

The caveat to the above argument is our assumption of a constant dissipation.  A non-uniform dissipation profile might be established if the shock weakens as it propagates, as expected, e.g., when a fast shell collides with a much thicker, slow shell.  In such a case the shape of the emitted spectrum can be  significantly altered, and in particular the thermal peak is expected to be shifted to much lower energies. 
This is illustrated in figure \ref{fig:spec4}, where the spectrum emitted from a shock moving at constant Lorentz factor, $\gamma_u=10$, is compared with that emitted from a decelerating shock with $\gamma_u(r_s)=10(r_0/r_s)^{1/2}$, where $r_0=10^{-2}r_{ph}$ is the initial deceleration radius of the shock.  The latter profile describes the dynamics of a blast wave propagating in a medium having a  density profile $n_b\propto r^{-2}$, and is naively expected at times longer than the crossing time of the reverse shock.   

In all the examples depicted in figures \ref{fig:spec1}-\ref{fig:spec3} the portion of the spectrum below the peak is much harder than that of a typical Band spectrum. This reflects a generic shortage in production of soft photons by sub-photospheric shocks.   The soft component may be produced subsequently by nonthermal processes at larger radii,  after the shock has emerged from the photosphere and became collisionless, or by multiple shock emission (see below).  On the other hand, the spectrum produced by a mildly relativistic shock ($\gamma_u\simlt 2$) just below the photosphere, as those exhibited in figure \ref{fig:spec2}, or by a decelerating relativistic shock that forms at a relatively large optical depth, as depicted in figure \ref{fig:spec4}, can nicely feet the thermal component observed in sources such as GRB090902B.  We emphasize that the efficiency of mildly relativistic, sub-photospheric shocks can be quite significant.  For instance, a shock propagating at $\gamma_u=2$ can dissipate up to 30\% of the bulk energy.  

A softer spectrum below the peak may be produced via multiple shock emission.   The energy dissipated behind a shock moving at a velocity $\beta_u$ scales as $\beta_u^2$ in the nonrelativistic regime (Eq. [\ref{erNR}]), and likewise the peak energy (Eq.  [\ref{kTsNR}]).  Thus, for a uniform distribution of shock velocities we expect $\nu E_\nu\propto\nu$.  In the ultra-relativistic regime the same argument yields $\nu E_\nu\propto\nu^{2}$.  Complications of this simple situation can be envisaged, but it is generally expected that the integrated emission from multiple shocks with a range of strengths, or a range of fromation radii, might lead to a spectrum that mimics a Band spectrum.  An elaborated treatment of multiple shock emission will be presented elsewhere. 
 
 \section{Conclusions}
We have analyzed the properties of subphotospheric shocks that form during the prompt phase of GRBs, and computed, under simplifying  assumptions, the time integrated spectrum emitted as the shock emerges from the photosphere.  We have shown that such shocks are mediated by Compton scattering of radiation advected into the shock by the upstream flow,  in difference from shocks that propagate in a cold medium, e.g., shock breakout in supernovae, where photons are generated inside the shock transition layer by free free emission.  We also argued that the scale of the shock, a few Thomson lengths, is vastly larger
than any kinetic scale involved, so that particle acceleration by the Fermi process is highly unlikely in such shocks.  The observed spectrum reflects the temperature profile downstream of the shock, with a possible contribution at the highest energies from the shock transition layer (BML11). 

Quite generally, in shocks that are dominated by photon advection the temperature just downstream of the shock transition layer, as measured in the shock frame, depends on the Lorentz factor of the upstream fluid and the ratio of photon and baryon densities advected into the shock.  In GRBs, the latter ratio is determined by the fireball parameters, specifically, the outflow power and the baryon load.   Since the immediate post shock region is photon starved, the temperature there is well in excess of the black body temperature.  Photon generation by double Compton and free-free emissions in the downstream flow tends to thermalize the radiation.  For typical GRB parameters, we find that the scale of the thermalization layer is about $10^4$ - $10^5$ Thomson depths, so that the radiation produced by sub-photospheric shocks that form in the coasting region is unlikely to reach full thermal equilibrium.   

The spectrum emitted by a single shocked shell that has reached the photosphere is a Wien spectrum, with a characteristic temperature roughly equals the immediate post shock temperature, modified by adiabatic cooling and thermalization.  The time integrated spectrum is the sum over all shells, and reflects the temperature profile behind the shock.   We have demonstrated that the emission from a single shock has a prominent thermal peak.  Such episodes can naturally account for the thermal component observed in GRB090902B and similar sources.  The location of the peak depends on the velocity profile of the shock; for a uniform dissipation profile (i.e., $\gamma_u=$ const) the thermal peak is located at  $h\nu_{peak}=3kT_{s,ob}$, with $T_{s,ob}$ given by Eq. (\ref{kTsR}) for RRMS and by Eq. (\ref{kTsNR}) for NRRMS.  The peak energy is significantly reduced in case of a decelerating shock (figure \ref{fig:spec4}). 

In general, the spectrum emitted by a single shock is very hard below the thermal peak, much harder than a typical Band spectrum.  However, we qualitatively demonstrated that multiple shock emission can in principle give rise to a much softer component below the peak, with $\nu F_\nu\propto\nu^s$, $1<s<2$, for mildly relativistic shocks with a uniform distribution of velocities.  Further work is needed to quantify the effect of multipole shock emission in more realistic situations, but the above reasoning suggests that multipole shock emission may mimic a Band spectrum. 

Unlike the case of an infinite planary shock, the structure of a shock that propagates in a relativistically expanding, nonuniform medium evolves with time. To be concrete, the shock width broadens, and the shocked fluid downstream becomes progressively more transparent as the shock moves outwards.  This raises the possibility that the trapped radiation downstream of the shock will re-interact with the shock as it approaches the photosphere, leading to a redistribution of the radiation energy.  Moreover, we find that a significant fraction of the total dissipation energy is contained inside the shock as it breaks out of the photosphere.  Consequently, the hard radiation produced inside the shock transition layer may significantly contribute to the high energy portion (above $3kT_{s,obs}$) of the observed spectrum.    A power law component extending up to about $\eta m_ec^2\simgt100$ MeV is conceivable.

\subsection*{Acknowledgments}
I thank Mitch Begelman, David Eichler  and Ehud Nakar for enlightening discussions. 
This research was supported by an ISF grant for the
Israeli center for high energy astrophysics.

\appendix
\section{\label{sec:tmprof}Computing the temperature profile}

The change in the photon density $n_r$ downstream of the shock, owing to free-free and double Compton emissions, is governed by the equation
\begin{equation}
\partial_\mu(n_ru^\mu)=\dot{n}_{ff}+\dot{n}_{DC},\label{app-cont-1}
\end{equation}
where $u^\mu$ is the 4-velocity of the shocked gas.  Assuming $n_e=n_p\equiv n_{bd}$ in the downstream flow,
the specific photon generation rate by Maxwellian e-p Bremsstrahlung emission can be expressed as  
\begin{equation}
\dot{n}_{ff}=\alpha_e\sigma_Tcn_{bd}^2(kT/m_ec^2)^{-1/2}\Lambda_{ff},\label{rate-ff}
\end{equation}
where $\Lambda_{ff}=10\Lambda_{ff1}=[-Ei(h\nu_{min}/m_ec^2)]g_{ff}$, $Ei(x)$ is the exponential integral 
of $x$, and $g_{ff}$ is the Gaunt factor.
  
Under the conditions considered here the Compton $y$ parameter exceeds unity for shocks that form at optical depth $\tau\simgt 10$.  For $y>>1$ the cutoff energy $h\nu_{min}$ is determined by the requirement that every newly generated photon can reach the thermal peak by scattering off the hot electrons  before getting re-absorbed. This yields $\Lambda_{ff}\simeq10$ at temperatures $kT_d$ of a few keV  (see, e.g., KBW09 and BML11 for details).   The rate of double Compton emission is
\begin{equation}
\dot{n}_{DC}=\frac{16}{\pi}\alpha_e\sigma_Tcn_{bd}(kT/m_ec^2)^{2}n_r\Lambda_{DC},\label{rate-DC}
\end{equation}
here $\Lambda_{DC}=\ln(kT/h\nu_{min})g_{DC}$.  For the ratio of the two rates we have, 
\begin{equation}
\frac{\dot{n}_{DC}}{\dot{n}_{ff}}=\frac{16}{\pi}(kT/m_ec^2)^{5/2}(n_r/n_{bd})(\Lambda_{DC}/\Lambda_{ff}).\label{app:ration-rates}
\end{equation}

Now, if the shock propagates in a stationary conical flow of a constant Lorentz factor, $u^0=c\eta>>1$, then Eq. (\ref{app-cont-1}) reduces to  
\begin{equation}
\frac{c}{r^2}\frac{d}{dr}(r^2 n_r\eta)=\dot{n}_{ff}+\dot{n}_{DC}.\label{app-con-spher}
\end{equation}
Dividing the last equation by $n_{bu}$, and recalling that $n_{bu} r^{2}=$ const and $n_{bd}=\sqrt{8}\gamma_u n_{bu}$, one obtains 
\begin{equation}
\frac{d}{d\tau}(n_r/n_{bu})=-8\alpha_e\gamma_u^2(kT/m_ec^2)^{-1/2}\Lambda_{ff}(1+  \dot{n}_{DC}/\dot{n}_{ff})\label{app-diff-ntil}
\end{equation}
in terms of the optical depth $d\tau=\sigma_Tn_{bu}dr/\eta$, where Eq. (\ref{rate-ff}) has been employed.  It can be readily shown that the last equation holds not only for a conical flow, but essentially for any geometry, provided the opening angle is sufficiently small.
Behind the shock the flow is adiabatic, so that the energy density of the radiation field satisfies $e_r\propto n_{bd}^{4/3}$.  If the density of the medium in which the shock propagates has a power law profile, $n_{bu}\propto r^{-\alpha}$, then $e_r\propto r^{-4\alpha/3}$ assuming $\gamma_u=$const.  For a shell that crossed the shock at a radius $r=r_s(t)$, where $r_s(t)$ is the shock trajectory, $e_{rs}$ is given by Eq. (\ref{erR}). At $r>r_s$ 
\begin{equation}
e_r=e_{rs}(r_s/r)^{4\alpha/3}=2n_{bu}\gamma_u^2m_pc^2(r_s/r)^{\alpha/3}.\label{app-er-addib}
\end{equation}
Now, on scales smaller than the thermalization length, the trapped radiation has a Wien spectrum with $n_r=e_r/3kT$.  We can therefore write
\begin{equation}
kT=\frac{2\gamma_u^2m_pc^2}{3(n_r/n_{bu})}(r_s/r)^{\alpha/3}.\label{app-temp-n}
\end{equation}
%We find it convenient to define a dimensionless temperature $\tilde{T}=T/T_s$.  
Substituting  Eq. (\ref{app-temp-n}) into (\ref{app-diff-ntil}) and using (\ref{app:ration-rates}), we arrive at
\begin{equation}
\frac{d\tilde{T}}{d\tau}-\frac{\tilde{\alpha}\tilde{T}}{3\tau}=A^R\tilde{T}^{3/2}[(\tau_s/\tau)^{\tilde{\alpha}/3}+\kappa^R \tilde{T}^{3/2}],\label{tmprof-diff-eq}
\end{equation}
where $\tilde{T}=T/T_s$, $\tilde{\alpha}=\alpha/(\alpha-1)$ ($\alpha>1$), and 
\begin{eqnarray}
A^R=12\alpha_e(m_e/m_p)\Lambda_{ff}(kT_s/m_ec^2)^{1/2},\label{AT-1}\\
\kappa^R=2\times10^3\gamma_u(kT_s/m_ec^2)^{3/2}.\label{kapp-1}
\end{eqnarray}
For our fiducial model, $L_{52}=\Gamma_0=R_6=1$, Eq. (\ref{kTsR}) yields $(kT_s/m_ec^2)=4(\gamma_u/\eta)$, and Eqs. (\ref{AT-1}) and (\ref{kapp-1}) reduce to 
\begin{eqnarray}
A^R=1.7\times10^{-4}\eta^{-1/2}\gamma_u^{-1/2}\Lambda_{ff1},\label{AR-2}\\
\kappa^R=1.6\times10^4\gamma_u^{5/2}\eta^{-3/2}.\label{kR-2}
\end{eqnarray}
Let $T_{ph}(\tau_s)$ denotes the temperature of a shell that crossed the shock at some radius $r_s<r_{ph}$ ($\tau=\tau_s>1$) as it reaches the photosphere ($\tau=1$).  Specifically, $T_{ph}(\tau_s)=T_s\tilde{T}(\tau=1)$, with $T_s$ given by Eq. (\ref{kTsR}).  For a given $\tau_s$,  $T_{ph}(\tau_s)$ is obtained by integrating Eq. (\ref{tmprof-diff-eq}) from $\tau=\tau_s$ to $\tau=1$, subject to the boundary condition $\tilde{T}(\tau=\tau_s)=1$.  Equation  (\ref{tmprof-diff-eq}) holds as long as $n_r<aT^3/k$ or equivalently $aT^4>e_{rs}(r_s/r)^{4\alpha/3}$.  

In the non-relativistic limit the temperature $\tilde{T}$ obeys Eq. (\ref{tmprof-diff-eq}) with $A^R$, $\kappa^R$ replaced by
\begin{eqnarray}
A^{NR}_T=3\times10^{-3}\beta_u^{-1}\eta^{-1/2}\Lambda_{ff1}(1+8\tilde{p}\beta_u^{-2})^{-3/2},\label{ATNR-1}\\
\kappa^{NR}=2\times10^4\beta_u^4\eta^{-3/2}(1+8\tilde{p}\beta_u^{-2})^{3/2}.\label{kappNR-1}
\end{eqnarray}

To compute the temperature profile behind an infinite planar RRMS, it is convenient to transform to the variable $\tau_d=\sigma_Tn_{bd} x^\prime$, where $x^\prime$ is the distance from the shock front, as measured in the frame of the shock.  The temperature then obeys the equation 
\begin{equation}
\beta_d\frac{d\tilde{T}}{d\tau_d}=-A^R\tilde{T}^{3/2}[1+\kappa^R \tilde{T}^{3/2}],\label{tmprof-planar-eq}
\end{equation}
with $A^R$ and $\kappa^R$ given by Eqs. (\ref{AR-2}) and (\ref{kR-2}), respectively.  The boundary condition in this case is $\tilde{T}(\tau_d=0)=1$.   Likewise, for a NRRMS, $\tilde{T}$ obeys the same equation with 
 $A^R$, $\kappa^R$ replaced by  $A^{NR}$, $\kappa^{NR}$ given above.

\section{\label{app:cond-coll}Upstream conditions for a collimated flow}
In \S \ref{sec:RMS-in-GRBs} we determined the conditions upstream of a sub-photospheric shock assuming radial geometry for the unshocked flow.   Here, we generalize the analysis to collimating flows. 
We consider a jet moving along the z-axis in a channel of varying cross-sectional radius $a(z)$.  For simplicity we adopt the profile 
$a(z)=a_0(z/z_0)^{\kappa}$.  The jet is characterized by a total power $L_j=n_bhm_pc^3\Gamma^2\pi a^2$, and mass flux $\dot{M}=n_bm_pc\Gamma\pi a^2$, where $h$ is the dimensionless enthalpy per baryon.  In the acceleration zone, where $h>>1$, the bulk Lorentz factor increases as $\Gamma=\Gamma_0(a/a_0)$.  The transition to the coasting regime, where $\Gamma=\eta\equiv L_j/\dot{M}c^2$, occurs at a distance $z_{coast}$ at which $a_{coast}\equiv a(z_{coast})=\eta a_0/\Gamma_0$:
\begin{equation}
z_{coast}=z_0(a_{coast}/a_0)^{1/\kappa}=z_0(\eta/\Gamma_0)^{1/\kappa},\label{rcoast-collim}
\end{equation}
and it is seen that the coasting radius of a collimated flow ($\kappa<1$) is larger than that of a radial flow by a factor $(\eta/\Gamma_0)^{(1-\kappa)/\kappa}$.

The optical depth at a distance $z$ from the origin can be expressed as 
\begin{equation}
\tau(z)\simeq\sigma_Tn_b z\Gamma^{-1}=\frac{\sigma_TL_jz}{m_bc^3\Gamma^3\pi a^2}=\frac{\sigma_TL_{iso}}{4\pi m_bc^3z\Gamma^3}=\frac{\eta_c^4z_0}{\Gamma_0\Gamma^3z},
\end{equation}
where the isotropic equivalent power is defined as $L_{iso}=4L_jz^2/a^2$, and $\eta_c$ is given by Eq. (\ref{eta_c-def1}) with $R_0=z_0$.
Thus, in the coasting region, where $\Gamma=\eta$, the optical depth is given by Eq. (\ref{tau-GRB}) also for a collimating flow.  The photosphere is located at $z_{ph}=z_0\eta_c^4/(\Gamma_0\eta^3)$.

Let $\tilde{\eta}_c$ denotes the critical load for which $z_{coast}=z_{ph}$.  From the above result for $z_{ph}$ and Eq. (\ref{rcoast-collim}) we obtain 
\begin{equation}
\tilde{\eta}_c=\Gamma_0(\eta_c/\Gamma_0)^{4\kappa/(1+3\kappa)}.\label{eta_c-coll}
\end{equation}
For a radial flow ($\kappa=1$) Eq. (\ref{eta_c-coll}) reduces to $\tilde{\eta}_c=\eta_c$, as it should.  For a collimating flow $\tilde{\eta}_c<\eta_c$.  
The relative location of the photosphere is given by $z_{ph}/z_{coast}=(\tilde{\eta}_c/\eta)^{(3+1/\kappa)}$, thus the photosphere will be located in the coasting region provided $\eta<\tilde{\eta}_c$.  

The dimensionless entropy $\tilde{n}$, Eq. (\ref{N-1}), depends only on the conditions in the vicinity of the central engine, not on the structure of the flow.  Consequently, Equation (\ref{N}) remains valid also in the general case of a collimated flow.  However, the ratio of radiation pressure and rest mass energy density does depend on the structure of the flow.  For the outflow profile adopted above we have
\begin{equation}
\tilde{p}(\tau)\simeq \left(\frac{r_{coast}}{r_\tau}\right)^{2/3}
\simeq(\tau)^{2/3}(\eta/\tilde{\eta}_c)^{2(1-\kappa)/3\kappa}.\label{til(p)-app}
\end{equation}
This difference merely affect the condition for formation of subrelativistic shocks, viz. $\beta_u>4\tilde{p}/3$, not so much the structure of RRMS.

For illustration, we compare the above results with the numerical simulations reported in Lazzati et al. (2009).  They analyzed the propagation of a GRB jet through the envelope of a 16 solar mass Wolf-Rayet progenitor star.  For numerical convenience the jet is injected at a distance of $R_{inj}=10^9$ cm 
with a total power of $L_j=5.3\times10^{50}$ erg/s, opening angle $\theta_0=10^\circ$, initial Lorentz factor $\Gamma_0=5$, and dimensionless enthalpy $h_0=80$ for which $\eta=h_0\Gamma_0=400$.  Fig 2 in their paper indicates that the transition to the coasting regime occurs at  $r_{coast}\simeq 10^{11}$ cm, thus it seems that in this simulation the core jet is nearly radial above the injection radius $R_{inj}$.
For the above values we find an isotropic equivalent luminosity at the injection point of $L_{iso}\simeq 7\times10^{52}$ erg/s.  It is difficult to predict what would the structure of the flow if injected at a realistic radius.  
If we naively interpolate the solution down to the central engine, at $R_0=10^6R_6$ cm, with $R_6\simeq1$ for a 3 solar mass black hole, we obtain $\eta_c=3\times10^3R_6^{-1/4}$.   This yields a photospheric radius of $r_{ph}=(\eta_c^4/\eta^3)R_0\simeq 1.2\times10^{12}$ cm, independent of $R_0$, in agreement with their numerical result. For the channel profile adopted above, a coasting radius of $r_{coast}\sim10^{11}$ cm corresponds to a collimating jet with $\kappa=0.6$.  Near the photosphere, where enhanced dissipation by collimation shocks takes place (see figure 2 in Lazzati et al. 2009), 
Eq. (\ref{til(p)-app}) gives $\tilde{p}\simeq0.1$.  Thus, NRRMS can form there provided $\beta_s>0.4$.  The spectra emitted by such shocks are well represented by Fig 3. For RRMS our model provides a reasonable description even if form below the coasting radius, at larger optical depths.

%%%%%%%%%%%%
\begin{figure}[ht]
\centering
\includegraphics[width=12cm]{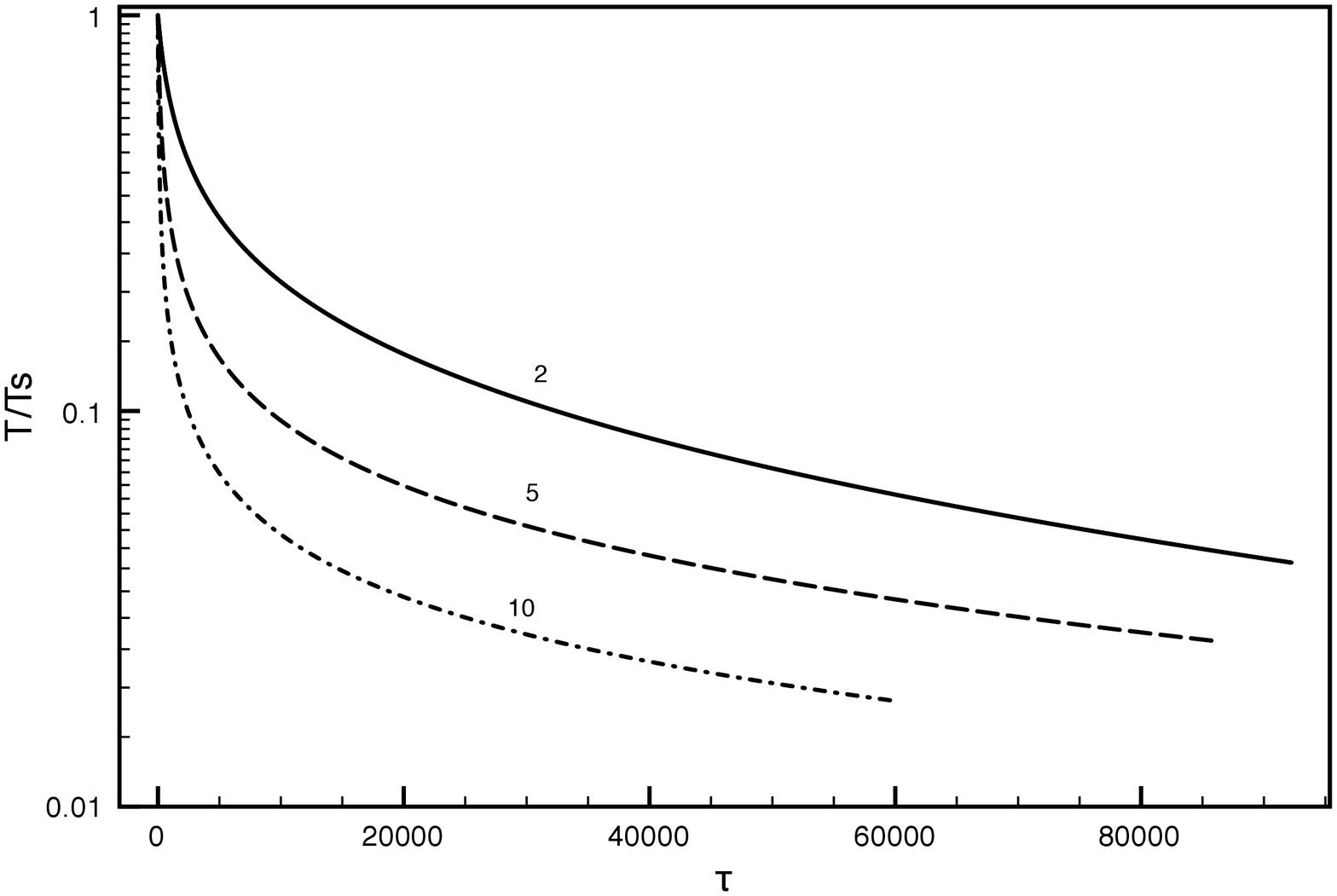}
\caption{\label{fig:tmprof}  Temperature profile in the thermalization layer of an infinite planar shock, for an upstream density $n_{bu}=10^{15}$ cm$^{-3}$, and different values of the shock Lorentz factor $\gamma_u$ (indicated by the numbers that label the curves).  }
\end{figure}

\begin{figure}[ht]
\centering
\includegraphics[width=12cm]{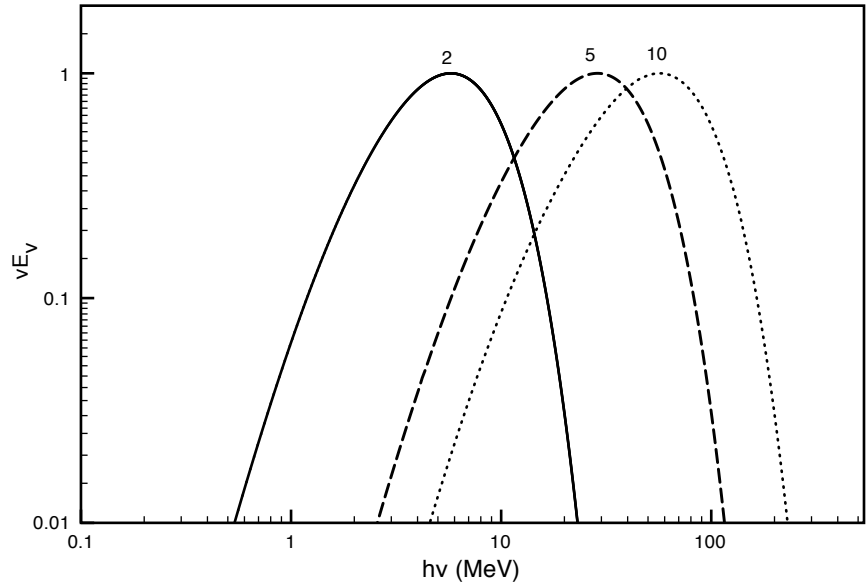}
\caption{\label{fig:spec1}  Time integrated spectral energy distribution for $\eta=200$, $\tau_0=10$, $\alpha=2$ (conical flow), and different values of the shock Lorentz factor $\gamma_u$ (indicated by the numbers that label the curves).  }
\end{figure}

\begin{figure}[ht]
\centering
\includegraphics[width=12cm]{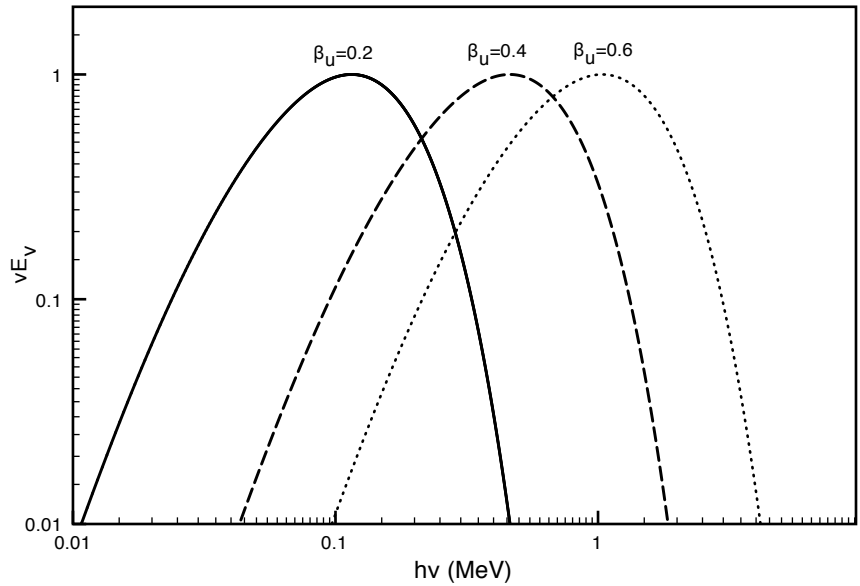}
\caption{\label{fig:spec2} Same as figure \ref{fig:spec1} for NRRMS with different shock velocities $\beta_u$.}
 \end{figure}

\begin{figure}[ht]
\centering
\includegraphics[width=12cm]{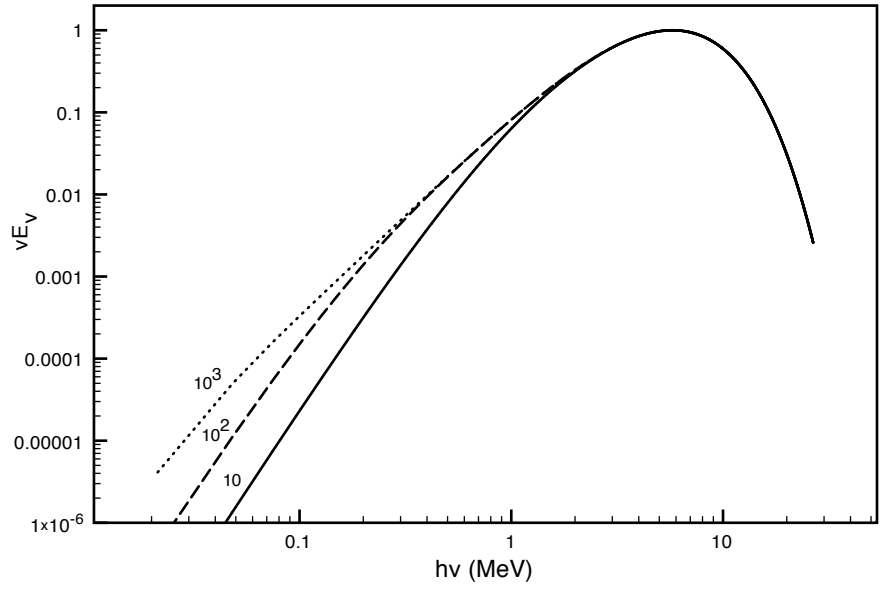}
\caption{\label{fig:spec3} Same as figure \ref{fig:spec1} for $\eta=600$, $\gamma_u=2$, and different values of $\tau_0$, as indicated.  The effect of adiabatic cooing is clearly seen at large values of $\tau_0$.}
 \end{figure}

\begin{figure}[ht]
\centering
\includegraphics[width=12cm]{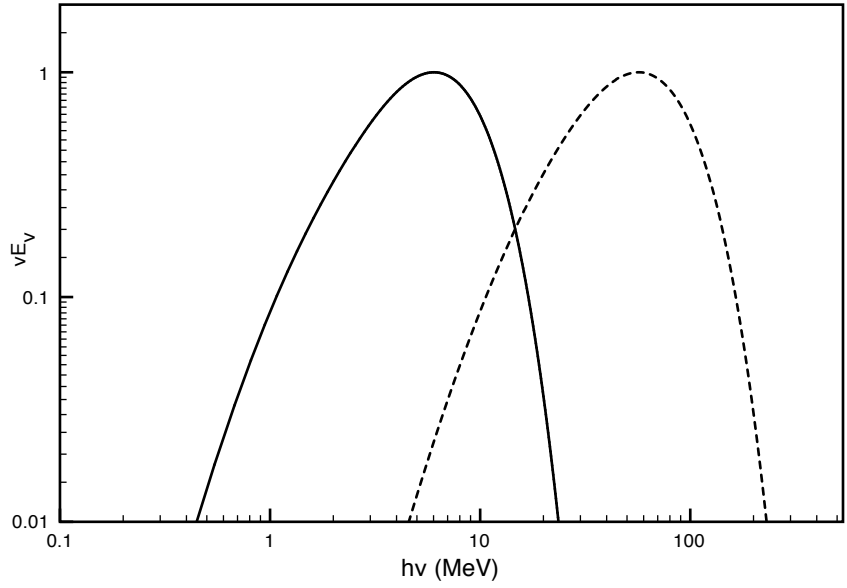}
\caption{\label{fig:spec4}The solid line delineates the time integrated SED produced by a decelerating shock having a Lorentz factor profile $\gamma_u(\tau_s)=10(\tau_s/\tau_0)^{1/2}$, with $\tau_0=r_{ph}/r_0=100$.   The unshocked flow in this example is taken to be conical ($\alpha=2$) with $\eta=200$.  The spectrum produced by a shock  moving at a constant Lorentz factor, $\gamma_u=10$, is displayed for a comparison (dashed line). }
\end{figure}

\end{document}